\documentstyle[11pt,moriond,epsfig]{article}

\def\be{\begin{equation}}
\def\ee{\end{equation}}
\def\bea{\begin{eqnarray}}
\def\eea{\end{eqnarray}}

\newcommand{\LL}{\ell^+ \ell^-}
\begin{document}
\vspace*{4cm}
\title{SEARCH FOR NON-SM HIGGSES AT LEP}

\author{ AURA ROSCA }

\address{Institut f\"ur Physik, Invalidenstr. 110,\\
D-10115 Berlin, Germany}

\maketitle\abstracts{
The four LEP experiments, ALEPH, DELPHI, L3 and OPAL,
have searched for Higgs bosons predicted by a large number
of extensions of the Standard Model. Flavor independent searches are presented
for the h$^{0}$Z$^{0}$ process in which the h$^{0}$ decays hadronically.
Search results are also presented for fermiophobic Higgs bosons,
invisibly decaying Higgs bosons, charged Higgs bosons and
the neutral Higgs bosons in the MSSM.}

\section{Introduction}

Since we have no experimental evidence for the Standard Model (SM) Higgs particle,
we must test more complicated scenarios beyond the SM.
Several extensions of the SM introduce additional Higgs doublets and singlets \cite{th}. 
In this work
we present results on the 2HDM Higgses. The 2HDM is a minimal
extension of the SM, in the sense that only one additional doublet is introduced in
the theory. After the symmetry is broken the Higgs spectrum consists of five particles:
two CP-even neutral scalars (h$^{0}$ and H$^{0}$), a
CP-odd neutral scalar (A$^{0}$) and a charged pair (H$^{\pm}$).
The properties of the Higgs bosons are defined by
six free parameters: four Higgs
masses, the ratio of the vacuum expectation values of the two fields,
$\tan \beta$, and the mixing angle in the neutral CP-even sector, $\alpha$. 
The Higgs couplings with the bosons and fermions control Higgs
production and decays. Particularly, the Higgs couplings with the
fermions must be limited, in order to suppress FCNCs (Flavor Changing
Neutral Currents) at the tree level. There are exactly two possibilities
for the Higgs boson to couple to the fermions: only one doublet
couples with fermions, and the other does not couple to quarks or leptons, and this 
determines the structure of the 2HDM of Type I; or one doublet couples to
down-type fermions, and the second doublet couples with the up-type
fermions, and this determines the structure of the 2HDM of Type II.

\section{Search for the Higgs bosons beyond the SM}
\subsection{Flavor independent searches}\label{subsec:flavor}

\begin{figure}[t]
\begin{minipage}{.46 \linewidth}
    \centering\epsfig{figure=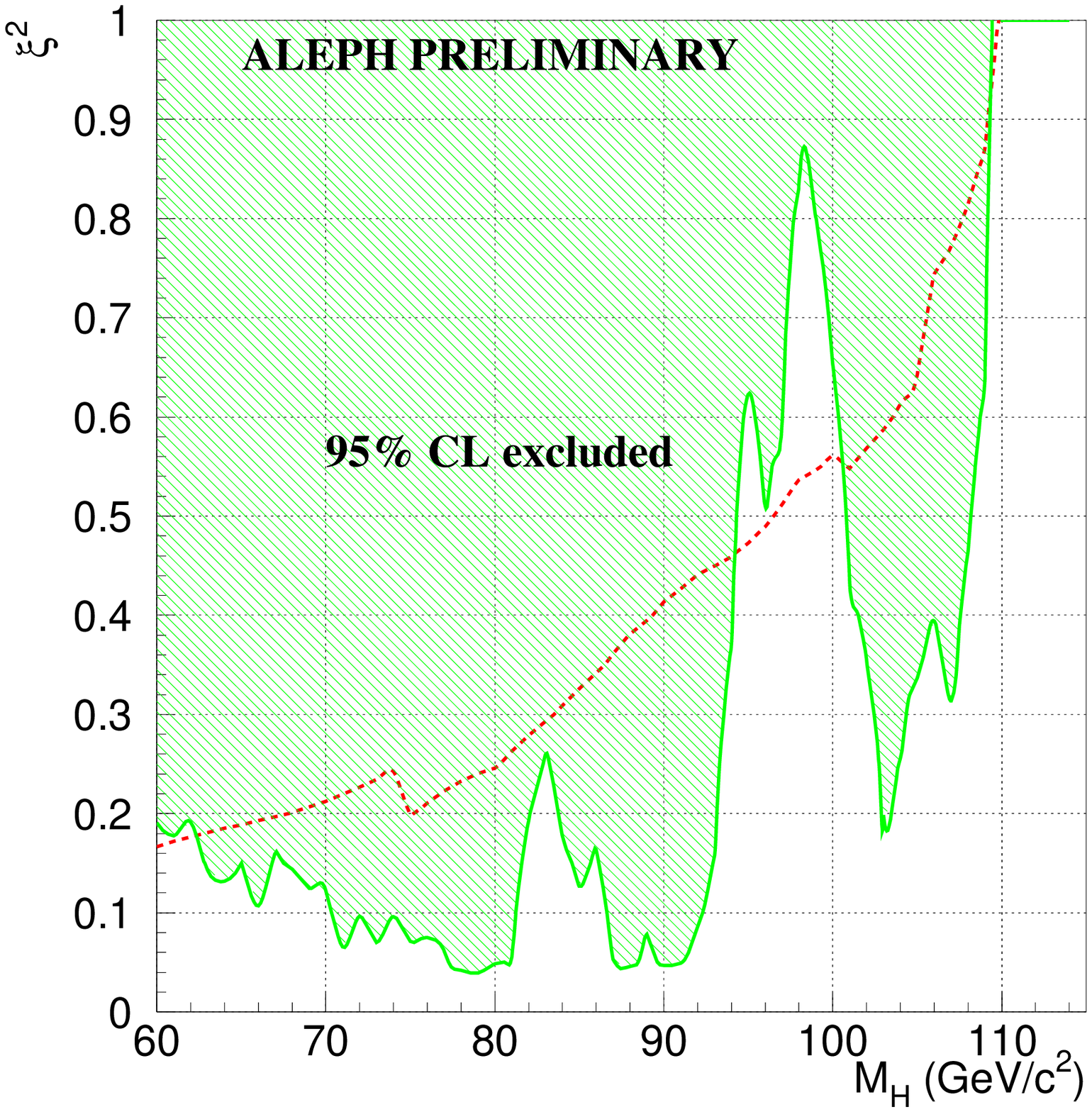,width=6.1cm}
    \caption{The expected (dashed line) and observed (hatched) 
95$\%$ C.L. excluded $\xi^{2}$ as a
function of the Higgs boson mass.} \label{flavor}
\end{minipage}\hfill
\begin{minipage}{.46 \linewidth}
    \centering\epsfig{figure=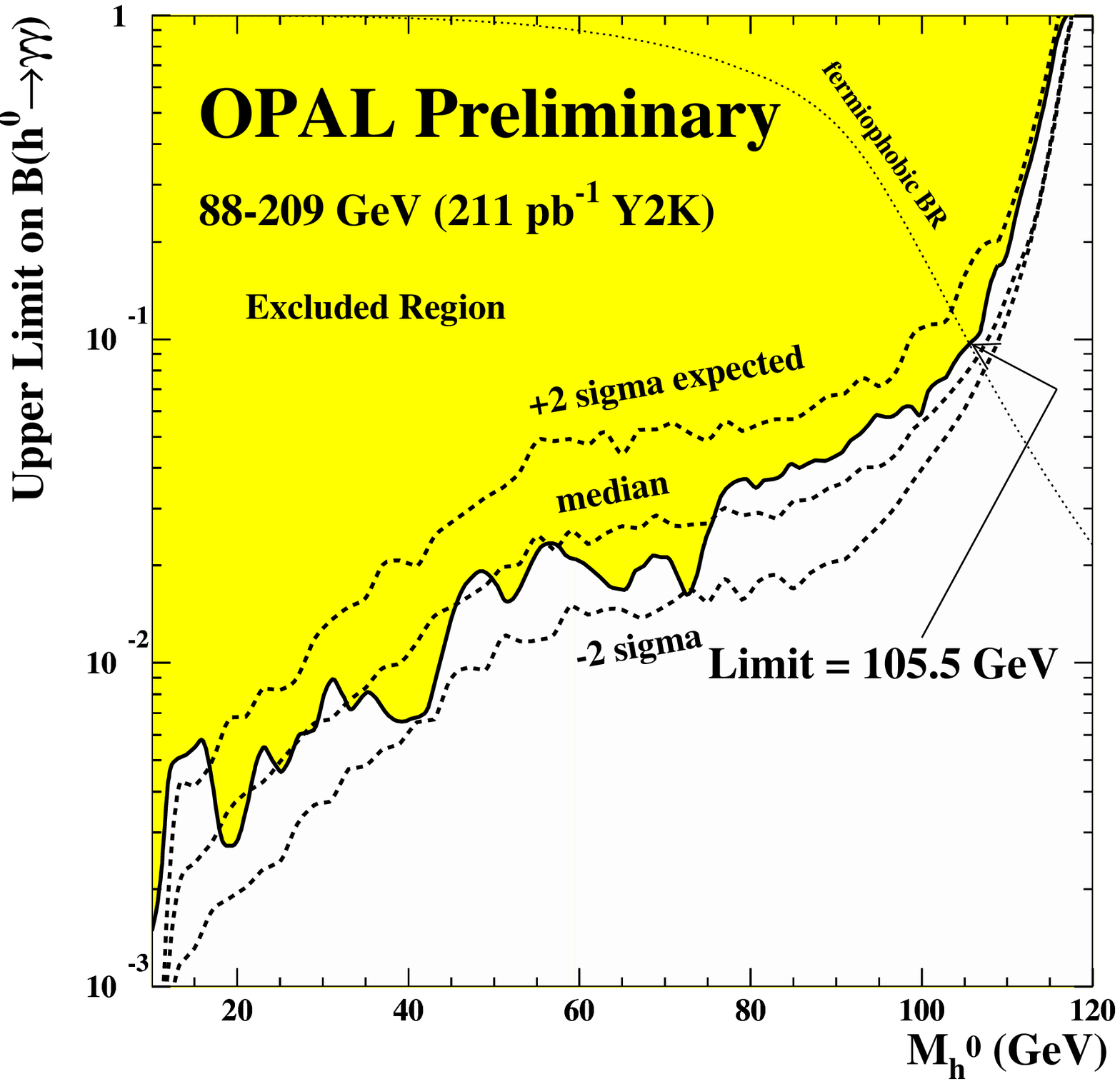,width=6.1cm}
    \caption{The expected (dashed line) and observed (shaded) 95$\%$ C.L. upper limit 
on the BR(h$\to \gamma \gamma$) 
assuming SM production cross section.}
\label{phobic}
\end{minipage}
\end{figure}

For certain regions of the parameter space the Higgs boson decay into a
pair of $\mbox{b}\bar{\mbox{b}}$ quarks can be suppressed with
respect to the SM values, while decays to charm quark or gluon pairs are enhanced. 
Therefore, flavor independent searches have been performed. 
The experiments have updated their search for a scalar boson which is produced in association
with the Z$^{0}$ and which decays hadronically with a BR = 1. These analyses
use the same topologies as for the SM Higgs boson search, but do not
make use of the b-tag.
In the absence of a signal, the results are usually given in terms of excluded 
regions in the plane
$m_{\rm h}$ versus a SM cross section scaling factor, $\xi^{2}$.
As an example we show in  Fig.~\ref{flavor}
the results obtained by the
ALEPH collaboration combining energies up to 209 GeV \cite{1}.
A Higgs boson produced with the SM cross section and decaying hadronically 
with a BR = 1 is ruled out by ALEPH up to the mass of 109.4 GeV. 
Similar results have been obtained by the OPAL
collaboration \cite{2}. 

\subsection{Fermiophobic decays}

Fermiophobic decays due to suppressed couplings between the Higgs bosons
and the fermions can appear in several non-minimal models, as for example 
2HDM of Type I. 
At masses below 90 GeV the Higgs boson decays dominantly into $\gamma \gamma$,
while at higher masses it decays mostly into WW$^{*}$ and ZZ$^{*}$.
The four LEP experiments searched for a photonically decaying Higgs boson
produced in association with a Z$^{0}$ and leading to the following final states:
$\gamma \gamma \mbox{q}\bar{\mbox{q}}$,
$\gamma \gamma \nu\bar{\nu}$,
$\gamma \gamma \LL$.
Since no indication for a signal was found in the data, 
the negative search result is given in terms of upper limits
of the BR$(\mbox{h}^{0}\to \gamma \gamma)$, assuming a SM production cross section
for the Higgs boson. The limits are shown in Fig.~\ref{phobic}
for the OPAL collaboration \cite{2}. 
In the fermiophobic model, the observed mass
limit is 105.5 GeV.
Similar results have been obtained by the other
collaborations \cite{3}.

\subsection{Invisible decays}

Invisible Higgs boson decays can appear in different models as MSSM, majoron
models, models with extra-dimensions, etc. At LEP energies it can be produced
via the Higgs-Strahlung mechanism, $\mbox{e}^{+}\mbox{e}^{-}\to 
\mbox{h}^{0}\mbox{Z}^{0}$, followed by
the decay of $\mbox{h}^{0}$ into undetectable final states. The signatures for such
processes are either 2 jets and missing energy in the event, or a lepton pair
and missing energy. The analyses make use of the kinematic constraint 
$m_{\rm f\bar{\rm f}} \approx m_{\rm Z}$. 
No excess with respect to the SM is observed in the data.
The results are expressed as excluded regions in the
plane $m_{\rm h}$ versus $\xi^{2}$, with $\xi^{2}$ defined as the ratio of the
Higgs boson production cross section times the branching fraction for the invisible Higgs decay,
normalized to the SM cross section.
For $\xi^{2}$=1 the lower limit on the mass of an
invisibly decaying Higgs boson is set by ALEPH at 114.1~GeV~\cite{4}. Similar results have
been obtained by the other collaborations  \cite{2}$^{,} $\cite{5}.

\subsection{Charged Higgs Bosons}

\begin{figure}[t]
\begin{minipage}{.46 \linewidth}
    \centering\epsfig{figure=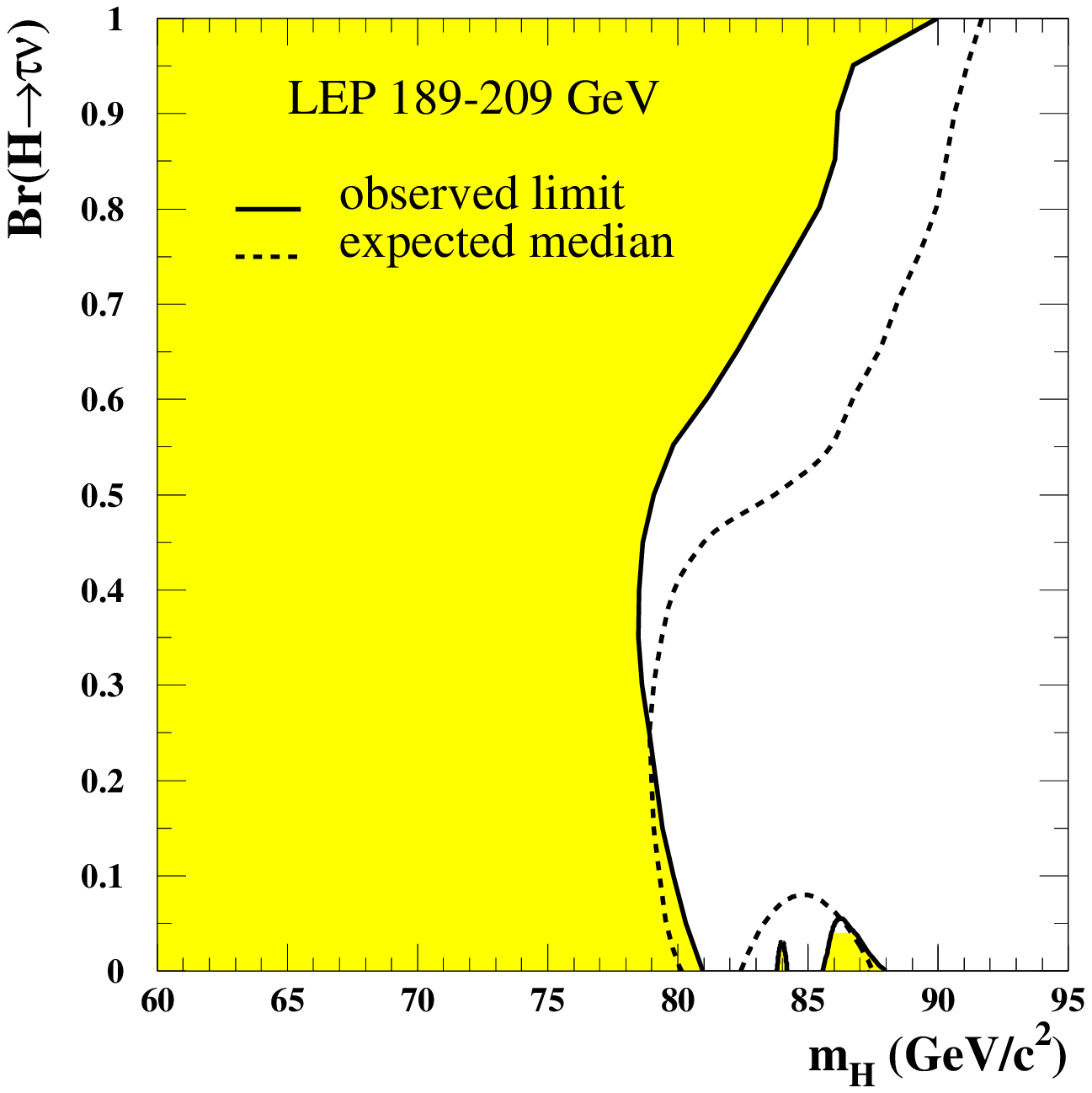,width=6.1cm}
    \caption{The expected (dashed line) and observed (full) 95$\%$ C.L. bounds on
$m_{\rm H^{\pm}}$ as a function of the branching ratio 
BR$(\mbox{H}^{+}\to\tau^{+}\nu_{\tau})$.
The shaded area is excluded at 95$\%$ C.L.} \label{chargedII}
\end{minipage}\hfill
\begin{minipage}{.46 \linewidth}
    \centering\epsfig{figure=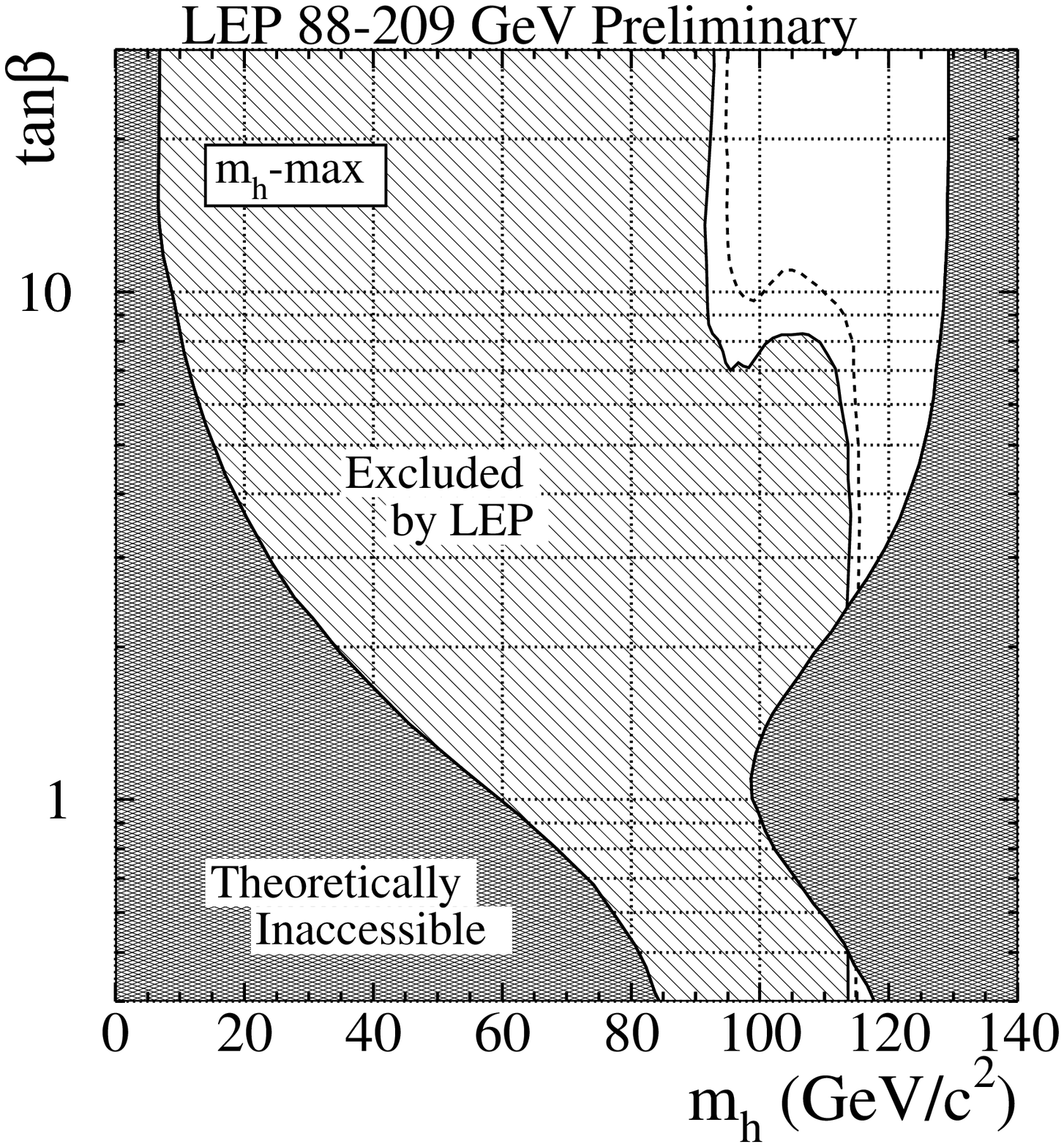,width=6.1cm}
    \caption{The MSSM 95$\%$ C.L. exclusion for the $m_{\rm h}$-max scenario. The excluded (hatched) and
theoretically forbidden (shaded) regions as a function of the MSSM parameters are shown. The expected
(dashed line) bound is also present.}
\label{mssm}
\end{minipage}
\end{figure}

Charged Higgs bosons are predicted in 2HDMs. At LEP energies, they are produced
in pairs, with a cross section which depends only on the Higgs boson mass. The
branching ratios are not predicted by the model. Possible decay modes are $\mbox{q}\bar{\mbox{q}}$,
$\tau \nu_{\tau}$ and W$^{*}$A$^{0}$. In 2HDM of Type II it is assumed that 
the first two fermionic decay modes are the only possible ones. The searches
cover all the possible final states: four jet final state, the semileptonic
and fully leptonic final states. The mass limits are expressed as a function of
BR$(\mbox{H}^{+} \to \tau ^{+} \nu_{\tau})$. The main background is from the process 
$\mbox{e}^{+}\mbox{e}^{-}\to \mbox{W}^{+}\mbox{W}^{-}$ which sets the scale for the limits in the hadronic
and semileptonic searches. Negative search results are shown in Fig.~\ref{chargedII}
combining the data from the four LEP experiments at energies up to 209 GeV \cite{6}.
The lowest mass limit obtained at BR$(\mbox{H}^{+}\to\tau^{+}\nu_{\tau})$=0.4 is 78.5 GeV, 
with a median expectation of 78.9 GeV. A small excluded "island" appears in Fig.~\ref{chargedII}
for BR$(\mbox{H}^{+}\to\tau^{+}\nu_{\tau})$=0, where the search 
sensitivity above the $\mbox{W}^{+}\mbox{W}^{-}$ background peak becomes sufficient to exclude this 
small region.  
In 2HDM of Type I, at high $\tan \beta$,
the decay mode $\mbox{H}^{\pm}\to \mbox{W}^{\pm *}A^{0}$ can become dominant, with a branching
ratio which depends on $\tan \beta$ and the mass difference between the
charged Higgs and the CP-odd neutral Higgs. 
The OPAL collaboration carries out such a search by considering new decay modes \cite{2}.
Results are expressed as excluded regions in the $(m_{\rm H^{\pm}},m_{\rm A})$
plane \cite{2}.

\subsection{Neutral Higgs Bosons in the MSSM}

The MSSM is a 2HDM of Type II. 
The Higgs boson production cross sections and the
branching ratios depend on the model parameters. 
We use a constrained model with only seven parameters:
the mass scale of the 
sfermions $M_{\rm SUSY}$, the gaugino mass parameter $M_{2}$, the supersymmetric
Higgs mass parameter $\mu$, the triliniar coupling of the Higgs boson to
the squarks A, the ratio of the vacuum expectation values of the two fields,
the mass of the CP-odd neutral scalar $m_{\rm A}$ and the gluino mass $m_{\rm g}$.
These parameters are further set to specific values which define three benchmark scans.
In the no-mixing scenario, the parameter which controls the mixing
in the stop sector, $X_{\rm t}=A-\mu \cos \beta$, is chosen to be zero. The other parameters 
are chosen to be $M_{\rm SUSY}=$ 1 TeV, $M_{2}=$ 200 GeV, $\mu=$ -200 GeV and 
$m_{\rm g}=$ 800 GeV and 
the scan is done over $0.4 < \tan \beta < 50$ and 4 GeV$ < m_{\rm A} < $1 TeV.
For the case of the maximal mixing in the stop sector, $X_{\rm t}=$2$M_{\rm SUSY}$ and the other 
parameters take the same values as before. The maximum mass value of the $\mbox{h}^{0}$ is
a function of $\tan \beta$. In the third scenario, called
the large-$\mu$ scenario, the parameters are chosen such as to emphasize points in the
parameter space where the Higgs decay mode into a $\mbox{b}\bar{\mbox{b}}$ pair is not dominant.
These values are: $M_{\rm SUSY}=$ 400 GeV, 
$M_{2}=$ 400 GeV, $\mu=$ -1 TeV and $m_{\rm g}=$ 200 GeV.

The two Higgs production mechanisms, the Higgs-Strahlung 
process, and the $\mbox{h}^{0}\mbox{A}^{0}$
associate production, vary in relative importance as a function of $\tan \beta$.
The $\mbox{h}^{0}\mbox{Z}^{0}$ 
production is dominant at low $\tan \beta$ and the MSSM phenomenology reduces to the
SM one, while the $\mbox{h}^{0}\mbox{A}^{0}$ 
production becomes dominant at high $\tan \beta$. The dominant decays of
the h$^{0}$ and A$^{0}$ Higgs bosons are into a pair of b-quarks and a pair of tau leptons. However, for
several choices of parameters, the decays h$^{0}\to$A$^{0}$A$^{0}$,  h$^{0}\to \mbox{c} \bar{\mbox{c}}$, 
h$^{0}$$\to$gg and 
h$^{0}\to$W$^{+}$W$^{-}$  can become~important. 
 
No evidence for a signal was found in the data. The negative results are presented as excluded regions in the
planes [$m_{\rm h},\tan \beta$], [$m_{\rm A},\tan \beta$],
[$m_{\rm h}, m_{\rm A}$] or [$m_{\rm H^{\pm}},\tan \beta$].
The 95$\%$ C.L. excluded region for the $m_{\rm h}$-max scenario is shown in Fig.~\ref{mssm}
combining the data from the four LEP experiments at energies up to 209 GeV \cite{7}. The lower limits on the
Higgs masses and the excluded $\tan \beta$ ranges for the $m_{\rm h}$-max and no-mixing scenarios
are summarized in Tab.~\ref{exp}.

\begin{table}[t]   
\caption{Limits on $m_{\rm h}$-max and no-mixing scenarios.
\label{tab:exp}}
\vspace{0.4cm}
\begin{center}
\begin{tabular}{|c|c|c|c|}   
\hline
Scenario  &
$m_{\rm h}$ limit (GeV) &
$m_{\rm A}$ limit (GeV) & 
Excluded $\tan \beta$\\ 
\hline
$m_{\rm h}$-max  &
91.0 &               
91.9 &   
0.5 $<$ $\tan \beta$ $<$ 2.4 \\
no-mixing  &
91.5 &               
92.2 &               
0.8 $<$ $\tan \beta$ $<$ 9.6 \\
\hline
\end{tabular}
\end{center}
\label{exp}
\end{table}

\section*{Acknowledgments}
I would like to thank the members of the four LEP collaborations and the LEP Higgs
working group for providing me plots and their help in preparing this talk.
I am grateful to Thomas Hebbeker for reading carefully this paper.

\end{document}